\documentclass[conference]{IEEEtran}

\ifCLASSINFOpdf

\else

\fi
\usepackage{enumerate}
\usepackage{graphicx}
\usepackage{comment}
\usepackage{tabularx}
\usepackage{amsmath}
\usepackage{cite}
\usepackage{breqn}
\usepackage{float}
\usepackage{balance}
\usepackage{mathtools}
\usepackage{amssymb}
\usepackage[font=small,skip=0pt]{caption}
\usepackage{amssymb}
\usepackage{float}
\usepackage{graphicx}
\usepackage{subfig}
\usepackage{gensymb}
\usepackage{lipsum} 
\usepackage{array}
\usepackage{color,soul}
\usepackage{optidef}
\usepackage{balance}
\IEEEoverridecommandlockouts
\usepackage[utf8]{inputenc}
\usepackage{cite}
\usepackage{amsmath,amssymb,amsfonts}
\usepackage{algorithmic}
\usepackage{graphicx}
\usepackage{textcomp}
\usepackage{xcolor}
\usepackage[export]{adjustbox}
\usepackage{url}
\usepackage{comment}
\providecommand{\e}[1]{\ensuremath{\times 10^{#1}}}

\makeatletter
 \let\old@ps@headings\ps@headings
 \let\old@ps@IEEEtitlepagestyle\ps@IEEEtitlepagestyle
 \def\confheader#1{%
 \def\ps@headings{%
 \old@ps@headings%
 \def\@oddhead{\strut\hfill#1\hfill\strut}%
 \def\@evenhead{\strut\hfill#1\hfill\strut}%
 }%
 \def\ps@IEEEtitlepagestyle{%
 \old@ps@IEEEtitlepagestyle%
 \def\@oddhead{\strut\hfill#1\hfill\strut}%
 \def\@evenhead{\strut\hfill#1\hfill\strut}%
 }%
 \ps@headings%
 }
 \makeatother

\confheader{%
 2017 Conference on Information and Communication Technology (CICT'17)
 }

 \usepackage[pscoord]{eso-pic}
\newcommand{\placetextbox}[3]{
 \setbox0=\hbox{#3}
 \AddToShipoutPictureFG*{ \put(\LenToUnit{#1\paperwidth},\LenToUnit{#2\paperheight}){\vtop{{\null}\makebox[0pt][c]{#3}}}
 }
 }
 \placetextbox{.5}{0.035}{\small{This paper has been accepted and presented in PESGM 2020 conference and is in the process for publication.}}

\begin{document}

\title{Data-Driven Power Electronic Converter Modeling for Low Inertia Power System Dynamic Studies
\thanks{This work is supported by the U.S. National Science Foundation under Grants Number MRI-1726964 and \textcolor{blue}{OAC-1924302}, the U.S. Department of Energy under Grant Number DE-SC0020281, and the SDSU Joint Research, Scholarship and Creative Activity Challenge Fund.}}

\author{\IEEEauthorblockN{Nischal~Guruwacharya$^\dagger$,  Niranjan~Bhujel, Ujjwol~Tamrakar, Manisha~Rauniyar, \\Sunil~Subedi, Sterling~E.~Berg, Timothy~M.~Hansen, and Reinaldo~Tonkoski}
\IEEEauthorblockA{ \textit{Department of Electrical Engineering and Computer Science} \\
\textit{South Dakota State University}\\
Brookings, South Dakota, USA 57007\\
Email: $^\dagger$nischal.guruwacharya@jacks.sdstate.edu}}

\maketitle
\thispagestyle{plain}
\pagestyle{plain}

\begin{abstract}
A significant amount of converter-based generation is being integrated into the bulk electric power grid to fulfill the future electric demand through renewable energy sources, such as wind and photovoltaic. The dynamics of converter systems in the overall stability of the power system can no longer be neglected as in the past. Numerous efforts have been made in the literature to derive detailed dynamic models, but using detailed models becomes complicated and computationally prohibitive in large system level studies. In this paper, we use a data-driven, black-box approach to model the dynamics of a power electronic converter. System identification tools are used to identify the dynamic models, while a power amplifier controlled by a real-time digital simulator is used to perturb and control the converter. A set of linear dynamic models for the converter are derived, which can be employed for system level studies of converter-dominated electric grids.
\end{abstract}


\section{Introduction}
With growing interest in renewable energy and batteries,  power electronic converters are becoming a crucial part of power distribution networks~\cite{bryne2018}. As the future energy demand is met by converter-based generation, models that accurately represent the interaction between the grid and the converters are essential. The response of these converter-based generations include fast-switching mechanisms that introduce faster and more stochastic dynamics compared to that of traditional power systems~\cite{Venkataramanan2004}. In the past, these dynamics were largely neglected as the percentage of converted-based generation was relatively low and the converters had a passive role as they were not actively contributing to voltage and frequency control of power systems. Neglecting the impacts of power electronics converters was possible as power system dynamics are largely dominated by large synchronous generators with well-defined models~\cite{kundur1994power}.   

To accurately model power electronic converters, one needs to have detailed knowledge of various aspects of a converter such as its physical topology, the complex models of the various voltage/current control loops, the models of the phase-locked-loop (PLL), the protection-scheme employed, etc. Even though the control architecture is known, these factors and control parameters vary significantly among manufacturers. This can lead to inaccurate modeling and simulation of the power system, resulting in erroneous results and analysis. Accurate models of converters can predict the instability of complex systems and verify the compatibility of components in a system~\cite{guarderas2019}. Such models are also essential for the proper design of controllers and protection systems.

Another factor that complicates modeling of the converter-based generation is to meet the requirements of grid interconnection and change in grid codes. For this, manufacturers can modify the control structure through a software/firmware update. For instance, as per IEEE 1547 standard, converters can actively participate in voltage and frequency support through advanced control functions\cite{ieee1547}. This adds another layer of complexity in modeling these converter systems. Black-box or data-driven models can be designed to solve the aforementioned issues. Recent accomplishments in data-driven modeling for inverters for system analysis are described in~\cite{guarderas2019,abbood2018,valdiviaTransient}. It is expected that power electronic converter dynamics will be different under different states of operation; however, there has been limited research on dynamic modeling of converters operating in different modes (e.g., dynamic real/reactive power support, ramp-rate control, voltage/frequency control)~\cite{guarderas2019}.
 
In this paper, a data-driven model of a commercial converter current output is obtained in response to changes in the voltage at the point of common coupling (PCC). The methodology used identifies reduced-order dynamics of the power electronic converter interfaced with the grid (or microgrid), which can then be used for system-level studies. The method does not require prior knowledge of the converter topology nor the implemented control approach. System identification tools, such as those provided by MATLAB \cite{MATLABtoolbox}, were used to develop these models. The models were developed by collecting input and output data from the actual inverter. 
These models can provide important information to the power system community to analyze the integration aspects of a large amount of converter based generation in power systems dynamics.

The paper is organized as follows: An overview of different data-driven black-box modeling of power electronic converters are presented in Section~\ref{state_of_the_art}. In Section~\ref{theory}, the theoretical background on the dynamics of power electronic systems and system identification is provided. The methodology used to obtain the dynamic model of a grid-connected inverter is illustrated in Section~\ref{methodology}. The results are presented in Section~\ref{results} followed by the main conclusions in Section~\ref{conclusions}.

\section{Data-Driven Modeling Approaches for Power Electronic Converters}

\label{state_of_the_art}
Power electronic converters for grid integration of renewable energy sources can consist of multiple and different cascaded and interconnected converters. The non-linearity of the switches used in these power electronic systems greatly increases the complexity of models. Detailed models of these converters are often used to perform accurate electromagnetic transient (EMT) simulations. Though accurate, the complexity of these models are prohibitive to be used for long-term and/or large-scale system studies~\cite{ninadGrid}. Furthermore, the parameters to accurately represent the exact dynamics are difficult to obtain and these methods cannot be employed without knowledge of the topology and control architecture used.  In cases where the switching dynamics are neglected, averaging techniques (specifically state-space modeling techniques) are often employed to derive small-signal transfer function models~\cite{middleebrook1982general}. However, depending on the analysis required even such state-space models may be computationally prohibitive~\cite{ninadGrid}.

Substantial efforts have been made to model the dynamics of such systems~\cite{statespace, valdiviaSystems, alskran, guarderas2019}. 
Detailed models can be developed using techniques such as average state-space modeling~\cite{statespace, alskran}. These models are very accurate and useful for component and converter level design~\cite{abbood2018, ninadGrid}, but they require detailed information about the converters~\cite{alskran}, which are often proprietary for commercial converters~\cite{valdiviaSystems}. Even if some of the internal parameters are known, the converter properties and dynamics may have a wide range of variation depending on load requirements, battery state-of-charge, and renewable energy availability. For these reasons, developing simplified models can be beneficial~\cite{ninadGrid}.

Data-driven modeling (or black-box modeling) is a useful method for modeling power electronic converters for system level studies~\cite{Ju_choi_system_identification, guarderas2019}. Black-box models can be developed with little to no information about the control or topology of a converter. As an additional benefit, black-box models usually require lower computational power compared to more detailed component level models~\cite{abbood2018}. Linear time-invariant (LTI) black-box models are often designed using regression analysis and curve fitting (described with more detail in ~\cite{Ljung}). Artificial neural networks (ANNs) can also be used to create black-box models~\cite{abbood2018}. Tools such as those provided by MATLAB's System Identification Toolbox \cite{MATLABtoolbox} are widely used for black-box modeling. The available modeling approaches range from simple linear models based on transfer functions to non-linear models using approaches such as the Hammerstein-Wiener model~\cite{Ljung},~\cite{Hammerstein_approach}. Black-box modeling of DC-DC converters has been widely explored in the literature~\cite{dcdcblackbox},~\cite{L_Arnedo} and more recently for DC-AC converters~\cite{statespace, valdiviaSystems, alskran, guarderas2019}. However, black-box models alone are not always accurate for a wide range operation~\cite{guarderas2019}. This variation over a wide operating range can be addressed by combining multiple models to cover the dynamics over the range-of-interest, for example in a polytopic structure~\cite{guarderas2019, valdiviaSystems}.

Data-driven modeling techniques from literature have mostly focused on the converters operating in standalone mode. These models may not be suitable when the converters actively interact with the grid and participate in grid ancillary services, such as providing voltage and frequency support. This is especially concerning for low-inertia power systems where power electronic converters will have a larger share of voltage and frequency control.

\section{Basic Concepts of Dynamic Modeling and System Identification }
\label{theory}

In this section, the dynamic modeling of a grid-connected inverter operating in current control mode is introduced. Among several power electronic converters, grid-connected inverters are widely used for interconnection of photovoltaic. So, we focus our discussion on this particular converter. This is followed by an introduction to basic concepts of system identification.

\subsection{Dynamic Modeling of Inverters}

\begin{figure}[h!]
            \centering
            \includegraphics[width=0.48\textwidth]{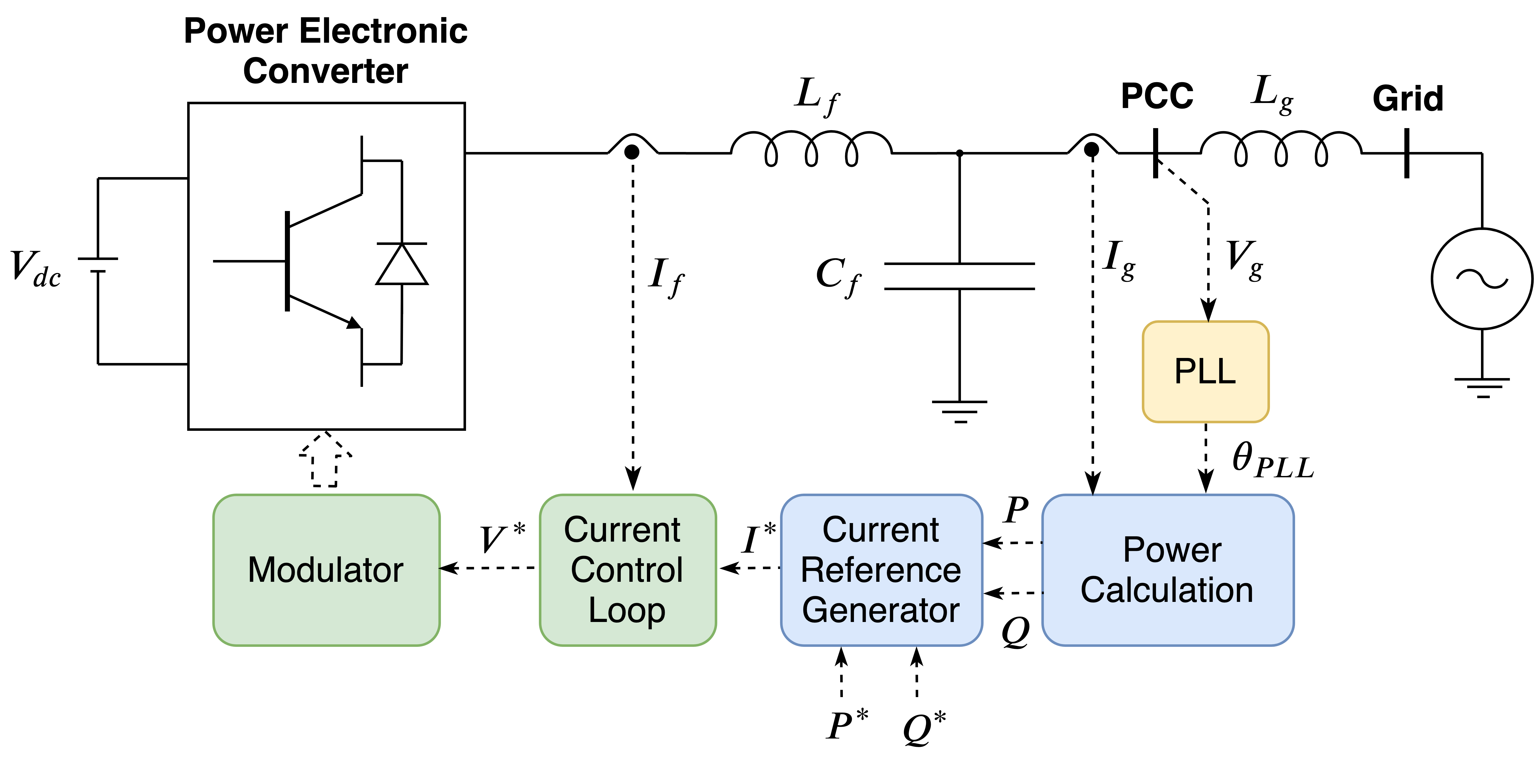}
            \caption{Schematic diagram illustrating the various components and control loops in a typical grid-connected inverter system.}
            \label{fig:inv_schematic}
        \end{figure}

A schematic of a grid-connected inverter system operating in current control mode is shown in Fig~\ref{fig:inv_schematic}. The inverter is connected to the electric grid through a low-pass filter with inductance $L_f$ and capacitance $C_f$; the inductance of the grid is represented by $L_g$. The inverter is being operated in the grid-following mode injecting the reference active and reactive power commands $P^*$ and $Q^*$ respectively. A PLL is used to track the phase-angle of the grid, $\theta_{PLL}$. Then a current controller (e.g., proportional-integral (PI)/ proportional-resonant (PR) controller) is employed to control the current being injected into the grid. 

The dynamics of this inverter system depends on various factors such as operation power level, DC voltage, parameters of the current controller, and the parameters of the PLL. The dynamic response of the grid current, for instance, depends largely on the control system being employed --- the design and controller gains will be different among various manufacturers. 

\subsection{System Identification of Power Electronic Systems}
\begin{figure}[h!]
        \centering
        \vspace{-8pt}
        \includegraphics[width=0.40\textwidth]{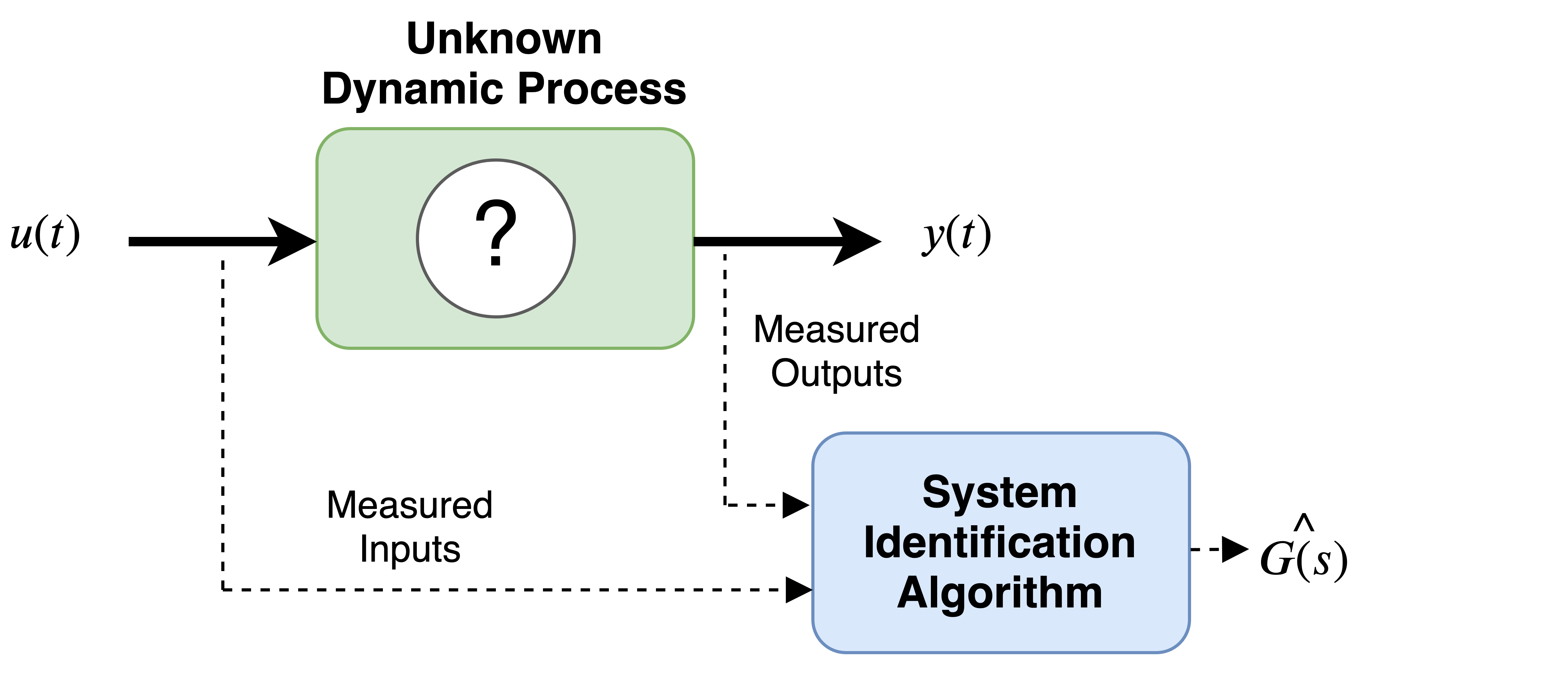}
        \caption{Basic concept of system identification. The system identification algorithm utilizes the input and output measurements to identify the unknown dynamic process.}
        \label{fig:identification}
    \end{figure}

System identification is a process to derive a mathematical model of an unknown system through observations of the input and corresponding output data. Using system identification tools, a mathematical model of a power electronic system that represents the dynamics of interest can be designed without knowledge of the underlying control structure and/or the control parameters. The dynamics of the converter will change based on different operating conditions. Several linear models for each operating condition can be developed and combined through a suitable mechanism~\cite{valdiviaSystems}. Fig.~\ref{fig:identification} illustrates the basic concepts of a system identification process. The input signal $u(t)$ and the output signal $y(t)$ are first measured from the unknown dynamic process to be identified. The dataset is then fed into a system identification algorithm which typically minimizes a defined cost-function to estimate the system model $\hat{G(s)}$.

The relationship between the input and output that can be defined as:
\begin{multline}
    \label{eqn:diffEqn}
    y(t)+a_1y(t-1)+ \dots +a_ny(t-n) =\\ b_1u(t-1) + \dots +b_mu(t-m)
\end{multline}

\noindent where $n$ and $m$ represent the number of poles and zeros of the system respectively. Similarly, $a_n$ and $b_m$ represents the parameters of the difference equation of (\ref{eqn:diffEqn}) or the coefficients of the equivalent transfer function. Then in general, a dynamic system can be represented as:
\begin{equation}
    \label{eqn:dynamic_sys}
    \hat{y}\left ( t\mid\theta\right )= \phi(t)^T\theta.
\end{equation}
In (\ref{eqn:dynamic_sys}), $\theta$ represents the set of the unknown parameters/coefficients of the system, and $\phi(t)$ represents the set of inputs $u(t)$ and outputs $y(t)$ of the dynamic system defined as follows:
\begin{equation}
    \label{eqn:poles_zeros}
    \theta = \left [a_1,\dots,a_n, b_1,\dots,b_m  \right ]^T
\end{equation}
\begin{equation}
    \label{eqn:input_output}
   \phi\left ( t \right ) = \left [ -y(t-1) \dots -y(t-n)~u(t-1) \dots u(t-m) \right ]^T 
\end{equation}

\noindent Now, if we define $Z^N$ as the set of known measurements and $N$ is overall input-output data in the time interval $1\leq t\leq N$:
\begin{equation}
    \label{eqn:meas_data}
    Z^N = \left \{ u(1), y(1), \dots, u(N), y(N) \right \}
\end{equation}

\noindent then the unknown parameters of the system, $\theta$, can be estimated by employing a least-squares method utilizing the following cost-function \cite{Ljung}:
\begin{equation}
    \label{eqn:cost1}
    \underset{\theta}{\text{minimize}} {\;}  V_N\left ( \theta, Z^N \right ).
\end{equation}
where
\begin{equation}
    \label{eqn:cost2}
    V_N\left ( \theta, Z^N\right ) = \frac{1}{N} \sum_{t=1}^{N} \left \| y(t) -\hat{y}(t\mid\theta) \right \|^2
\end{equation}

Based on the collected input-output data, a set of models with different numbers of poles and zeros can be fitted to the data. The fit of the model can be calculated using a metric such as the normalized root-mean-square-error (NRMSE) defined as~\cite{MATLABtoolbox}:
\begin{equation}
    \label{eqn:NRMSE}
    NRMSE = 1 - \frac{\left \| y(t) - \hat{y}(t\mid\theta) \right \|}{\left \| y(t) - \text{mean}{\;}{(\hat{y}(t|\theta))} \right \|}
\end{equation}

\noindent Furthermore, to compare different models based on the goodness of fit and complexity of the model the  Akaike's Final Prediction Error (FPE) can be used, defined as~\cite{MATLABtoolbox}:
\begin{equation}
    \label{eqn:FPE}
    FPE = \text{det} \left( \frac{1}{N} \sum_{t=1}^{N} \left ( e(t,\hat{\theta_N}) \right ) \left ( e(t,\hat{\theta_N}) \right )^T \right )\left ( \frac{1+\frac{d}{N}}{1-\frac{d}{N}} \right )
\end{equation}

\noindent where $e(t)$ represents the prediction errors and $d$ is the number of estimated parameters. A lower FPE represents a more accurate model of the system. 

\section{Methodology}
\label{methodology}
\begin{figure}[h!]
        \centering
        \includegraphics[width=0.45\textwidth]{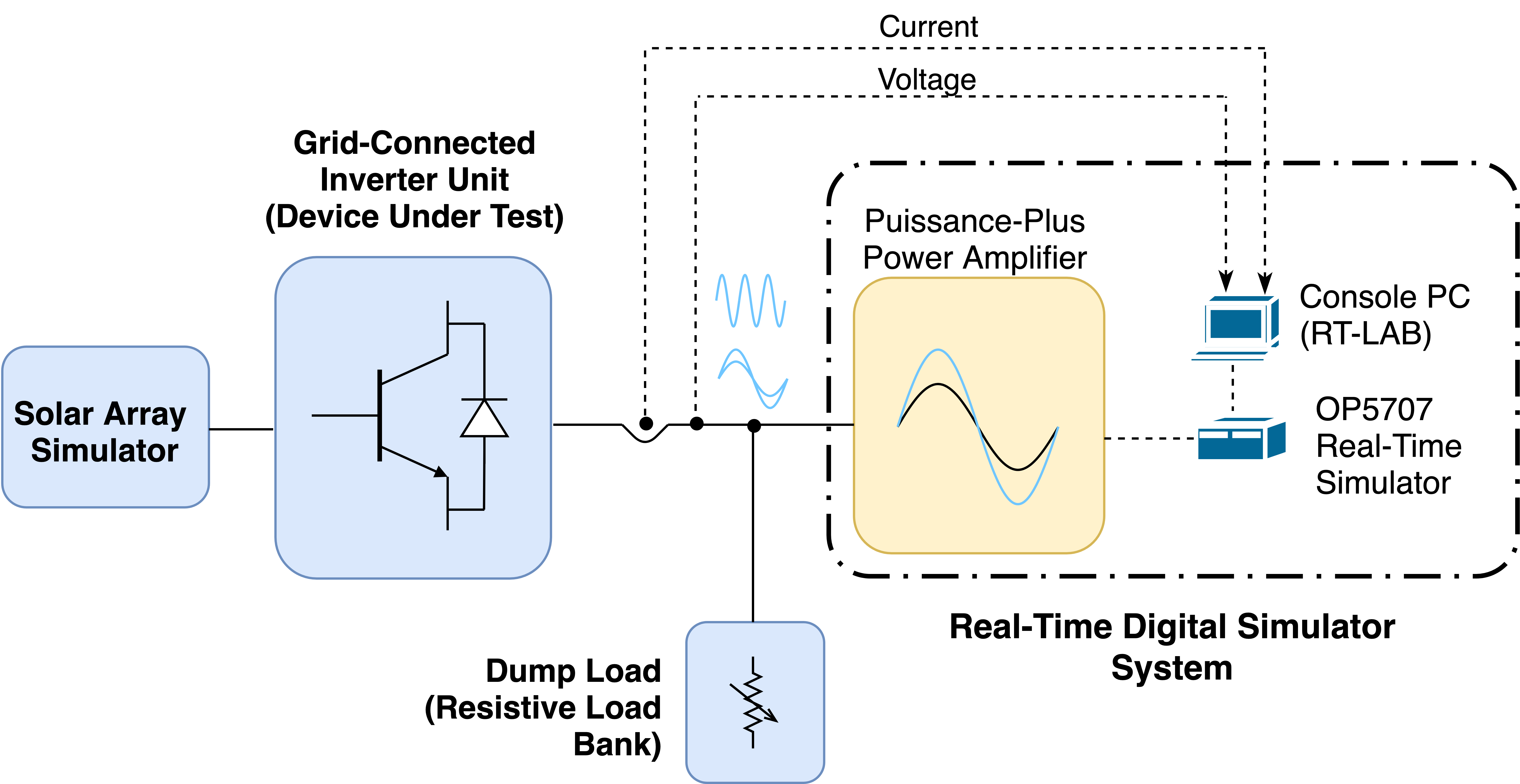}
        \vspace{10pt}
        \caption{Experimental setup for system identification. The grid-connected inverter is probed through a power amplifier unit controlled through an Opal-RT real-time simulator.}
        \label{fig:setup}
       \vspace{-5pt}
    \end{figure}

The experimental setup used for identifying the dynamics of a grid-connected inverter is shown in Fig.~\ref{fig:setup}. The device under test is a 700 W grid-connected inverter from SMA (Sunny Boy SB 700U) whose transfer function is to be determined. A solar array simulator (SAS) was used to emulate the DC output of a PV system and the test device is connected to an Opal-RT which consists of a OP5707 real-time simulator combined with a power amplifier from Puissance-Plus. In conjunction with the console PC, the real-time simulator and the power-amplifier unit can emulate grid voltage of varying output magnitude, phase, and frequency. A resistive dump-load is also connected to consume excess power that cannot be consumed by the power amplifier. The nameplate rating of the grid-connected inverter from SMA is given in Table~\ref{tab:Inverter_Rating}.

 \begin{table}[h]
     \centering
     \caption{SMA Grid-Connected Inverter's Nameplate Ratings (Sunny Boy SB 700U).}
     \begin{tabular}{|c|c|}
     \hline
        Nominal voltage  &  120 V\\
        \hline
        Voltage Range  &  106-132 V\\
        \hline
        Frequency Range & 59.3-60.5 Hz\\
        \hline
        Nominal frequency  &  60 Hz\\
        \hline
        MPPT range  &  75-200 Vdc\\
        \hline
        \end{tabular}
     \label{tab:Inverter_Rating}
 \end{table}

To determine the transfer function of the grid-connected inverter, the dynamic response of the inverter current is observed when there are perturbations in the grid voltage. This situation may be more common now as the grid becomes susceptible to overvoltage issues and as per IEEE 1547 standard the inverters can have voltage ride-through capabilities. To emulate this scenario, the amplitude of the power amplifier's output voltage is varied through the Opal-RT real-time simulator. This emulates over/under-voltage conditions in the grid. The corresponding output voltage at the PCC and the current supplied by the inverter are logged through the Opal-RT system. The voltage and current measurements are fed into the console PC, where the dynamics of the device under test will be identified using the System Identification Toolbox available in MATLAB/Simulink to implement the system identification technique described in the prior section. 

Assuming, $b_m$ and $a_n$ are the coefficients of the numerator and denominator respectively, the transfer function to be identified is:
\begin{equation}
    \label{eqn:transfer_function}
    \hat{G}(s) = \frac{\Delta i_{inv}(s)}{\Delta v_g(s)} = \frac{b_ms^m + b_{m-1}s^{m-1} + \dots + b_0 }{a_ns^n + a_{n-1}s^{n-1} + \dots + a_0} 
\end{equation}
where $\Delta i_{inv}(s)$ and $\Delta v_{g}(s)$ are deviation in current and voltage from normal operating point. Using this transfer function, the poles and zeros of the system can be identified.
\section{Results and Analysis}
 \label{results}
\begin{figure}[h]
    \centering
    \vspace{-8pt}
    \includegraphics[width=0.53\textwidth]{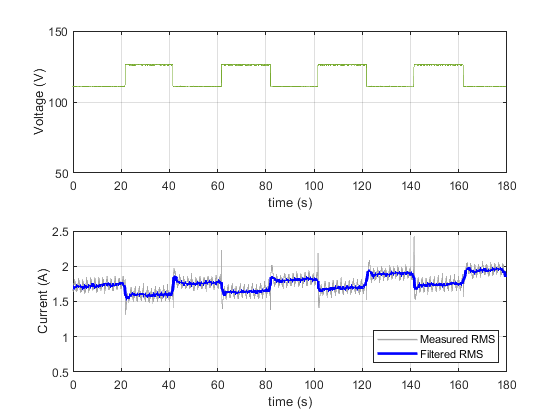}
    \caption{Response of inverter output current with step change in grid voltage.}
    \label{fig:collected_data}
\end{figure}

\begin{figure}[h]
    \centering
    \vspace{-12pt}
    \includegraphics[width=0.53\textwidth]{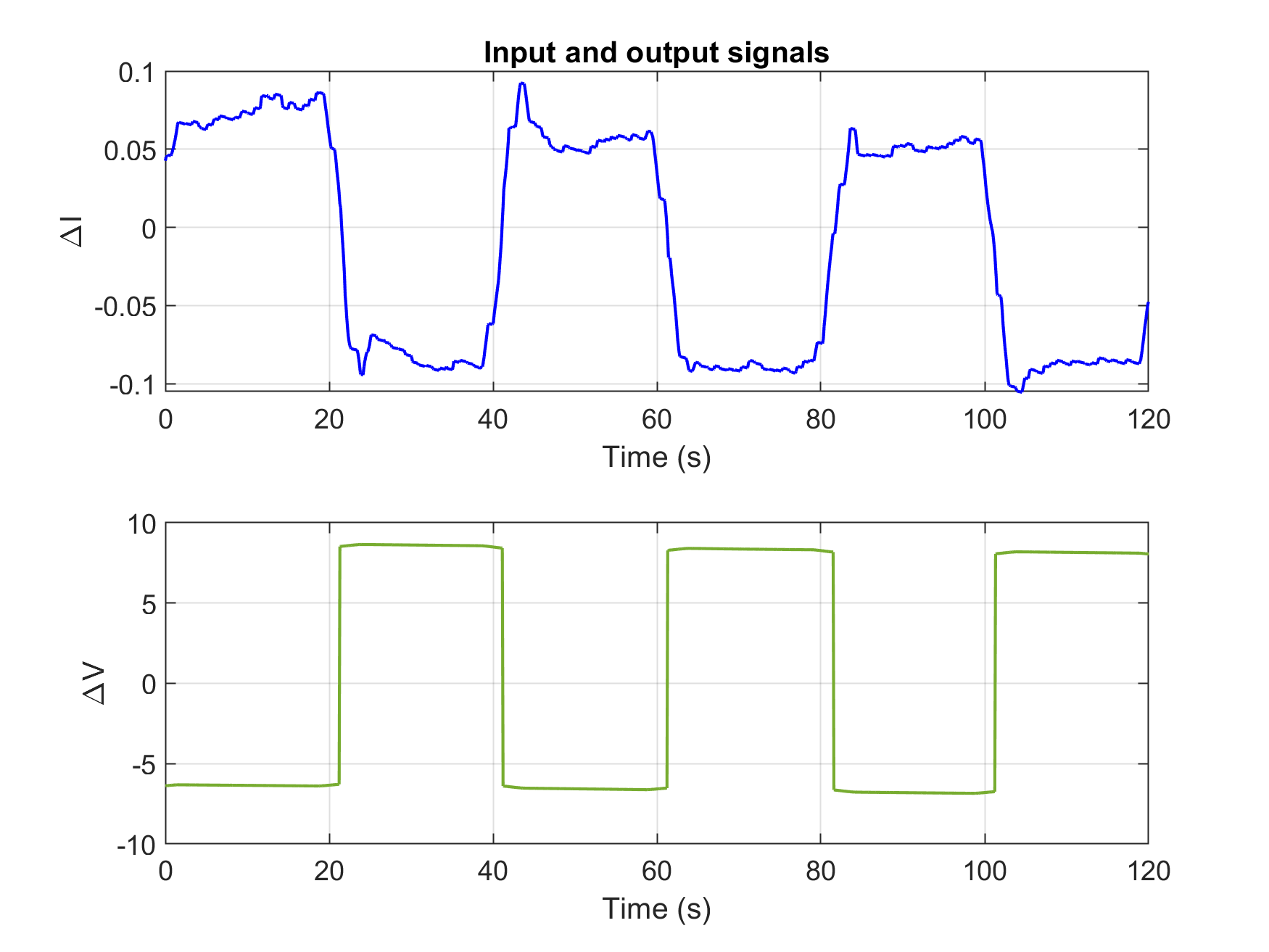}
    \caption{Training dataset obtained after pre-processing the measured current and voltage signals.}
    \label{fig:processed_data}
\end{figure}

 Figure~\ref{fig:collected_data} shows the response of the SMA inverter to the changes in the grid voltage. The root-mean-square (RMS) value of both the voltage and current signal is shown. A median filter was applied which works as a non-linear digital filter and smooths the array of sampled data, preserves edges while eliminating unwanted noise signals. The process of calculating the RMS value introduced a linear trend in the current readings. This makes the data unsuitable for system identification as this is an artifact from the pre-processing and not an actual part of the dynamics of concern. This linear trend is thus minimized through the detrend tool available within the System Identification Toolbox.  Furthermore, to get a more accurate model, mean of both current and voltage measurements are removed. This allows the focus of the identification to be on the actual fluctuations due to the perturbations rather than unwanted trends in the data. For cross-validation purposes, the dataset is split into a training set to compute the unknown poles and zeros and testing set to validate the derived model. The training dataset obtained after proper pre-processing is illustrated in Fig.~\ref{fig:processed_data}.

The possibility of getting a better fit through higher-order models was also explored. For this, a system with 3-poles and 1-zero; and 3-poles and 2-zeros were analyzed. Table~\ref{tab:summ_table} lists the various models that were fitted along with a metric that demonstrate the goodness of fit for training and testing data. A transfer function with 2-poles and 1-zero seems to provide the \textit{best} fit. The goodness of fit was highest for this case with both the testing and training dataset. Furthermore, the Akaike's FPE metric is also the least from this case compared to the other two cases. Increasing the number of poles from 2 to 3 in the second case slightly reduced the goodness of fit. Similarly, increasing both poles and zeros as in the third case slightly increases the goodness of fit against the testing dataset. Based on this analysis the following second-order transfer function was identified to be suitable:

\begin{equation}
    \label{eqn:tf_estimate}
    \hat{G(s)} = \frac{\Delta i_{inv}(s)}{\Delta v_g(s)} = \frac{-0.02113s - 9.334\e{-4}}{s^2 + 2.104s+0.1133} 
    \vspace{6pt}
\end{equation}

\begin{table}[h]
\centering
\caption{Summary of transfer function models identified through the System Identification Toolbox.}
\label{tab:summ_table}
\begin{tabular}{|c|c|c|c|c|}
\hline
\begin{tabular}[c]{@{}c@{}}Model \\ Order\end{tabular} & \begin{tabular}[c]{@{}c@{}}Model \\ Coefficients\end{tabular} & \begin{tabular}[c]{@{}c@{}}Fit to \\ Training\\ Data\end{tabular} & \begin{tabular}[c]{@{}c@{}}Fit to \\ Test\\ Data\end{tabular} & FPE \\\hline
\begin{tabular}[c]{@{}c@{}}$n$ = 2\\ $m$ = 1\end{tabular} & \begin{tabular}[c]{@{}c@{}}$b_1$ = -0.02113\\ $b_0$ = -9.334\e{-4}\\ $a_2$ = 1.000\\ $a_1$ = 2.104\\ $a_0$ = 0.1133\end{tabular} & 76.77\% & 74.24\% & 3.118\e{-4} \\ \hline
\begin{tabular}[c]{@{}c@{}}$n$ = 3\\ $m$ = 1\end{tabular} & \begin{tabular}[c]{@{}c@{}}$b_1$ = -0.06635\\ $b_0$ = 1.6\e{-4}\\ $a_3$ = 1.000\\ $a_2$ = 4.344\\ $a_1$ = 6.701 \\ $a_0$ = 0.012222\end{tabular} & 74.2\% & 72.45\% & 3.85\e{-4} \\ \hline
\begin{tabular}[c]{@{}c@{}}$n$ = 3\\ $m$ = 2\end{tabular} & \begin{tabular}[c]{@{}c@{}}$b_2$ = -0.02651\\ $b_1$ = -9.392\e{-3}\\ $b_0$ = -0.1478\\ $a_3$ = 1.000\\ $a_2$ = 3.024\\ $a_1$ = 6.661\\ $a_0$ = 14.69\end{tabular} & 76.17\% & 73.92\% & 3.28\e{-4} \\ \hline
\end{tabular}
\end{table}

\begin{figure}[h]
    \centering
    \vspace{-12pt}
    \includegraphics[width=0.45\textwidth]{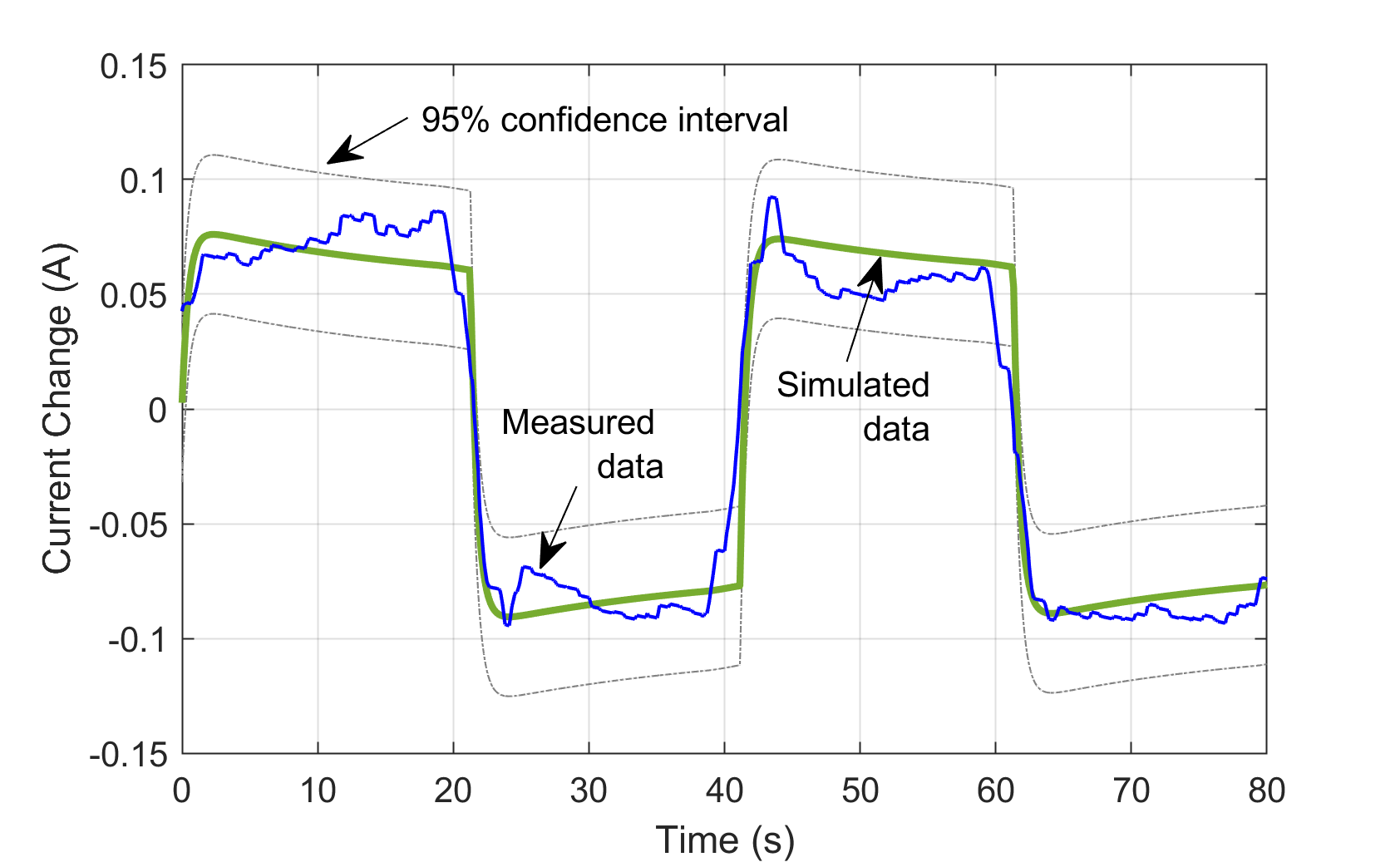}
    \caption{Measured versus simulated output of the fitted transfer function along with the 95\% confidence interval of the estimate.}
    \label{fig:meas_vs_sim}
\end{figure}

The performance of this model is also illustrated in Fig.~\ref{fig:meas_vs_sim} by comparing the simulated model against the measured data. The simulated model is the response that is computed based on the fitted model, using the test data as the input. Ideally, the simulated model should be very close to the measured data for a good model fit. The fit obtained in this case was 76.77\% which is slightly on the lower side. The dotted lines illustrate the 95\% confidence interval of the estimates. The confidence interval represents the range of output values having 95\% probability of being the true response of the system. 

\section{Conclusions and Future Work}
A data-driven, black-box model for a grid-connected inverter was developed in this paper. The voltage at the PCC of the inverter was perturbed through a power amplifier unit. Using the MATLAB's System Identification Toolbox, the logged voltage and current dataset were used to identify several transfer function models. The models were validated on a testing dataset and based on different metrics that measure the goodness of fit, a second-order model was identified to best fit the data. In the future, we intend to explore the dynamics of different inverters under several operating conditions or modes of operation. The different linear transfer function models will be combined through a statistical approach to derive a generalized non-linear model that captures the most significant dynamics of the inverter. These generalized non-linear models can be used to develop and analyze converter-dominated systems.

\label{conclusions}
\balance

\balance
\bibliographystyle{IEEEtran}
\vspace{-6pt}
\bibliography{references}

\end{document}